\documentclass[11pt]{article}
\usepackage{geometry}

 \usepackage{graphics}
 \usepackage{graphicx}
 \usepackage{epsfig}
 \usepackage[all]{xy}

 \usepackage{amssymb,amsmath,amsthm}
 \usepackage{latexsym, epsfig, psfrag,eepic,colordvi,bm}
 \usepackage{graphics,color}
 \usepackage[all]{xy}
 \usepackage{epstopdf}
 \usepackage{appendix}

 \usepackage{graphicx}
\usepackage{graphicx,epsfig}
\usepackage{algorithmic, algorithm}

\usepackage{caption,subcaption}

\newtheorem{theorem}{Theorem}[section]

\newtheorem{proposition}[theorem]{Proposition}

\newtheorem{definition}{Definition}[section]

\DeclareMathOperator{\Deg}{deg}
\newcommand{\tp}{$[{\bf t}, {\bf p}]$}

\title{Independent decisions collectively producing a long information dissemination path with a foreseen lower-bounded length \\in a network}
\author{Ricky X. F. Chen~\footnote{ORCID: 0000-0003-1061-3049}\\
	\small School of Mathematics, Hefei University of Technology\\[-0.8ex]
	\small Hefei, Anhui 230601, P.~R.~China\\[-0.8ex]
	\small\tt chenshu731@sina.com
}

\date{}

\begin{document}
\maketitle

{\bf\noindent Abstract.}
Our research problems can be understood with the following metaphor: In Facebook or Twitter, suppose Mike decides to send a message
to a friend Jack, and Jack next decides to pass the message to one of his own friends Mary, and the process continues until the current message holder could not find a friend who is not in the relaying path.
How to make the message live longer in the network with each individual's local decision? 
Can Mike foresee the length of the longest paths starting with himself in the network by only collecting information of native nature?
In contrast to similar network problems with respect to short paths, e.g.,
for explaining the famous Milgram's small world experiment, no nontrivial solutions have been proposed for the problems.

The two research problems are not completely the same and notably our approach yields solutions to both.
We discover node-specific numeric values which can be computed by only communicating identity-free degree derivatives
to network neighbors and for an arbitrary network node $v$ there exists a function determining a lower bound for the length of the longest paths starting with it based on its numeric values. Moreover, in the navigation process initiated at $v$, inspecting the numeric values of their neighbors,
the involved nodes can independently make their decisions eventually guaranteeing a path of length longer than the determined lower bound at $v$.
Numerical analyses demonstrate plausible performance of our approach of inferring certain global properties from local
information in complex networks.




\newpage

%%%
%%%%%%%%%%%%%%%%%%%%%%%%%%%%%%%%%%%%%%%%%%%%%%%%%%%%%%%%%%%%%%%%%%%%%%%%%%%%%%%%%%%%%%%%%%%%%
%%%
%\section{Introduction}
%%%
%%%%%%%%%%%%%%%%%%%%%%%%%%%%%%%%%%%%%%%%%%%%%%%%%%%%%%%%%%%%%%%%%%%%%%%%%%%%%%%%%%%%%%%%%%%%%
%%%
{\section{Introduction}\label{sec:introduction}}

{S}{tructural} properties of large-scale networks are of central importance for
understanding the formation principles of said networks as well as the dynamics associated to them,
and one of the key challenges is inferring properties of global nature from local information.
An important structural property that has been widely studied concerns paths in networks,
including determining the length of paths and searching paths, as it is of practical significance in information routing
and transportation, etc.
In some scenarios, short paths are desired, while in others, long paths are more relevant.
Here we are concerned with the latter.
Let us first introduce our research problems in a real-life metaphor as follows.
Suppose in Facebook or Twitter or any other social network, Mike decides to send a message
to one of his friends Jack, and Jack next decides to pass the message to one of his own friends Mary, and the process continues until the present message holder could not find a friend who is not involved in the relaying path.
How to make the message live longer in the network, i.e., to be relayed as many hops as possible, with each individual's local decision? 
% How can Mike deduce a nontrivial estimate for the length of the longest paths starting with himself in the network by only collecting information of native nature?
Another close problem that we ask is if Mike can foresee the length of the longest paths starting with himself in the network by only collecting information of native nature.
These problems may arise in other networks and have alternative interpretations as well. For instance,
in self-organized network (SON)-based sensor networks or alike, in certain situations,
a message (including a physical object) generated at an agent (i.e., a node or vertex) may simply need relaying as many hops as possible without a deterministic
destination.
%Optimization in a specific application is beyond our discussion here.
In these network scenarios, there are no powerful central controllers having all detailed structural information
due to security or cost or other reasons. Hence, agents have to independently make their own decisions based on local information they can acquire.
While the content of local information is generally left open and deserves exploration, here we propose an explicit assumption that an agent is only allowed to communicate identity-free numeric value(s) derived from node degrees in order for better evaluating relevant solutions. 

For a comparison purpose, we remark that finding short paths from a node to another node in a network with local information was first
studied by Kleinberg~\cite{kl00a,kleinberg} in order for explaining Milgram's small world experiment.
Through this experiment, the ``small-world" nature or the ``six degree of separation" principle is very well disseminated.
But another fact implied in the experiment was not really noticed at the beginning and pointed out later by Kleinberg.
That is, not only exist short paths between any pair of strangers, but people could also collectively find a short path connecting them using
only local information.
Kleinberg proposed a lattice network model where he can show in certain case an approach of searching a short path locally
does exist.
Many subsequent works on finding short paths with local information have been discussed, for instance, for power-law networks in~\cite{LRAB} and scale-free networks in~\cite{KYHJ}.
Note that finding the shortest paths from any fixed node to any other nodes using global structural information can be easily solved, e.g., by the well-known breadth first search algorithm or Dijkstra's algorithm.

Searching a longest path in a network is widely known to be NP-hard as it is a generalization of the famous Hamiltonian problem. 
%It is even shown that for any $\epsilon <1$, the problem of finding a path of length $n-n^{\epsilon}$ in an $n$-vertex Hamiltonian graph is NP-hard~\cite{KMR}.
 Algorithms for finding paths with a fixed length or a prescribed performance ratio, can be found in~\cite{ AYZ, zl, BHus,koli} and references therein.
A number of lower bounds for the length of the longest paths or cycles, in particular in terms of the minimum degree or the number of edges of the graph in question, have been obtained, e.g.,~in Dirac~\cite{dirac}, Erd\H{o}s and Gallai~\cite{erdos-gallai}, and Alon~\cite{alon}, etc.
But, these results are generally of global optimization nature, not node specific, and do not seem to be applicable to our problems.
There are some natural navigation strategies though, for instance, randomly picking neighbors or randomly picking neighbors of maximal degree.
We shall employ these random-based strategies as benchmarks for evaluating our solution to the navigation problem.

We emphasize that the two long-path problems are not completely the same: the message sourse is not necessarily aware of the overall length of the relaying path even though the implemented navigation strategy may be fairly good (e.g., a random-based strategy may produce a long path but
the message source has no way to know that), and that a node can derive a good estimate of the length of the longest paths starting with itself does not guarantee an applicable ``one-way" relaying strategy attaining the estimated length.
Notably, our approach yields solutions to both.

In order for providing a solution to the long-path problems of an arbitrary network regardless of the application
scenarios, our first contribution is a very general new framework for analyzing network structures, called \tp-separate chain decomposition of networks. From a \tp-separate chain of a network, every node there receives a rank. We next prove that the ranks for all nodes can be computed via searching fixed points (steady states) in certain dynamical systems on the network, with the worst-case $O(nm)$ time complexity for a network of $n$ nodes and $m$ edges. The computation can be carried out distributively at nodes which are only allowed to communicate degree derivatives with their
neighbors.
As such, the ranks are essentially derivate from only local information.
Finally, the rank of a node is shown to be the kind of local information that can help the node determine the length of the longest paths starting with or containing itself, and formulate a strategy to relay messages.
   Although there is unfortunately no theoretical justification due to its hard nature at present, evaluations on a great number of typical real-world networks, Barab\'{a}si-Albert type random networks and Erd\H{o}s-R\'{e}nyi random networks demonstrate that our navigation strategy generally outperforms the natural strategies of random navigation.

\section{Results}

\subsection{Local information proposed}\label{sec2}

Let $G=(V,E)$ be a simple graph, where $V$ is the set of nodes (or vertices) and $E$ is its edge set.
We write $H\leq G$ if $H$ is a subgraph of $G$.
In the following, if not explicitly specified otherwise, a network of $n$ nodes has $[n]=\{1,2,\ldots, n\}$ as its node set. We sometimes write $v\in G$ for $v\in V(G)$.

\begin{definition}
Let $\mathbb{Z}$ be the set of integers and
$
{\bf t}=(t_1, t_2, \ldots, t_n),\, {\bf p}=(p_1, p_2, \ldots, p_n)\in \mathbb{Z}^n.
$
Suppose $\mathcal{C}: G_0, G_1 ,  G_2 ,  \ldots , G_m$ is a chain of subgraphs of a network $G$ of $n$ nodes, where $G_0=G$, and $G_i$ is a vertex-induced subgraph of $G_{i-1}$ for $1\leq i \leq m$. We call the chain $\mathcal{C}$ a $[{\bf t},{\bf p}]$-separate chain if for any $0\leq i \leq m$, any vertex $v\in G_i$ has
\begin{itemize}
\item[(a)]  at least $\max\{0,i+p_v\}$ neighbors in $G_i$, and 
\item[(b)] at least $i$ neighbors in $G_j$ where $j=\max \{0, i+t_v\}$.
\end{itemize}
\end{definition}
 We call the number $m$ the length of the chain $\mathcal{C}$, and denote it by $L(\mathcal{C})=m$. The number $k$ such that $G_k\neq \varnothing$ while $G_{k+1}=\varnothing$ or $k+1>m$ is called the size of the chain $\mathcal{C}$, and denoted by $|\mathcal{C}|=k$. We sometimes write the chain $\mathcal{C}$ as $G=G_0 \geq G_1 \geq  G_2 \geq \cdots \geq G_m$.

These chains aim at providing new insight into ``positions" of nodes in the global structure of the network $G$. Namely, we intend to characterize the internal mutual connection of a ``community" in the global structure and the ability of the nodes there reaching the outside of the community.

For a node $v$ of $G$, we denote by $\Deg_G(v)$ (or $\Deg(v)$ if $G$ is clear) the degree of $v$ in $G$.
For some ${\bf t}, {\bf p}\in \mathbb{Z}^n$, there may be no \tp-separate chains for $G$ at all.
The following proposition gives a sufficient and necessary condition for the existence of \tp-separate chains.

\begin{proposition}[Existence]\label{prop:existance}
	Let $G=(V, E)$ and ${\bf t}, {\bf p}\in \mathbb{Z}^n$.
	Then, there exists a \tp-separate chain of $G$ if and only if $p_v \leq \Deg(v)$ for any $v\in V$.
\end{proposition}

The proof of the above proposition can be found in the Supplementary Information (SI).
Hereafter, we assume $p_v \leq \Deg(v)$ for any node $v$ unless they are
explicitly specified.

\begin{definition}[Maximal chains]
	A \tp-separate chain $\mathcal{C}$ of $G$: $G_0 \geq  G_1  \geq \cdots \geq G_m$ is called maximal if there does not exist a \tp-separate chain $\mathcal{C}': G'_0\geq G'_1  \geq \cdots \geq G'_{m'}$ satisfying either (i) $|\mathcal{C}'|> |\mathcal{C}|$, or (ii) $|\mathcal{C}'|= |\mathcal{C}|$ and for some $ 1\leq i \leq |\mathcal{C}|$, $G_i < G'_i$.
\end{definition}

A concept in the literature that can be equivalently formulated as a special instance in our \tp-chain framework is
the classical $k$-core of a network.
The $k$-core of a network $G$ is the maximal subgraph of $G$
where any vertex has degree at least $k$~\cite{k-core}.
It is left to the reader to verify that the chain consisting of the $k$-cores of $G$ is actually a maximal $[{\bf 0, 0}]$-separate chain of $G$, where ${\bf 0}=(0,0,\ldots, 0)$.
There are many applications of
the $k$-core decompositions to real-world network problems, see~\cite{kcore1, kcore3, np-core,nc-core, core-like} for instance.
A discussion on the distributed computation of $k$-core can be found in~\cite{MPM} and later in~\cite{nc-core}.
Note that merging two \tp-chains entry-wisely gives another \tp-chain.
Therefore, it is not hard to show the following crutial uniqueness result.

\begin{proposition}[Uniqueness]
	Let $G=(V, E)$ and ${\bf t}, {\bf p}\in \mathbb{Z}^n$ where $p_v \leq \Deg(v)$ for any $v\in V$.
	Then, there exists a unique maximal \tp-separate chain of $G$.
\end{proposition}

In the maximal \tp-separate chain, if a vertex $v\in G_k$ and $v\notin G_{k+1}$, then we denote $C_{\bf t, p}(v)=k$.
Apparently, $G_k$ is just the vertex-induced subgraph of $G$ by the set of vertices $v$ with $C_{{\bf t,p}}(v)=k$. 

{\bf Distributive computation of $C_{\bf t, p}(v)$.}
Inspecting the definition of \tp-separate chains, it is not clear how to obtain such chains,
not to mention distributively.
In fact, we do not know any efficient approach for obtaining all chains at present.
However, we do find an approach towards obtaining the maximal \tp-separate chain if exists.
The approach is based on a novel connection discovered between the maximal \tp-separate chain
of a graph $G$ and a certain fixed point of some discrete dynamical system on $G$.

A discrete dynamical system over a network~\cite{linear2,linear1,kauf,vonn,wolf,rei5} is concerned with
the dynamics generated when nodes in the network update their states following a system update schedule and their respective rules.
Von Neumann's cellular automata~\cite{vonn} are such dynamical systems.

%Let $G=(V,E)$ be a graph with $V=[n]$.
Let $P$ denote a finite set of states a node in $G$ may have and let $x_i$ denote the state of node $i$.
A function $f_{i}$ specifies how node $i$ updates its state $x_{i}$ based on the states of the neighbors of $i$
(and itself) in $G$.
A fair update schedule is
an infinite sequence of subsets of vertices $W=W_1W_2\cdots$, where
for any $k\geq 1$ and any $1\leq i \leq n$, there exists $l>k$ such that $i\in W_l$.
Suppose the initial system state at time $t=0$ is ${\bf x}^{(0)}=(x^{(0)}_1, x^{(0)}_2, \ldots, x^{(0)}_n)$.
For $j>0$, the system state ${\bf x}^{(j)}$
at time $t=j$ follows from that the nodes contained in $W_j$ update their states via
applying their respective functions to the states of their respective neighbors in ${\bf x}^{(j-1)}$
while the states of the nodes not contained in $W_j$ stay unchanged.
We denote this dynamical system by $[G,f,W]$, and we denote by
$[G,f,W]^{(j)}({\bf x})$ the system state at time $t=j$ when the system starts
from the initial state ${\bf x}$ at time $t=0$.

({\bf Fixed point}) In a dynamical system $[G,f,W]$, we say a system state ${\bf x}$ is
reaching a fixed point (or steady state) ${\bf z}$ if there exists $k> 0$ such that for any $j>k$,
$$
[G,f,W]^{(j)}({\bf x})=[G,f,W]^{(k)}({\bf x})={\bf z}.
$$

({\bf $[{\bf t},{\bf p}]$-system}) Let $G$ be a graph on $[n]$.
Here we are interested in a particular class of systems, where
each vertex of $G$ can have a state from the set $[n]$, and the function $f_v$ at a vertex $v$ returns the maximum $k$ such that there are at least $k$ of the neighbors of $v$ with (state) values at least $k+t_v$ while at least $\max\{0, k+p_v\}$ of them with values at least $k$.
For example, suppose $t_v=-2$ and $p_v=-1$, and $\{2,4,4,5,3\}$
is the (multi)set of values of the neighbors of $v$. Then,
$f_v$ returns $4$.
We call such a system the $[{\bf t},{\bf p}]$-system on $G$.
It turns out that these systems are useful in computing maximal chains,
and $W$ can be an arbitrary fair update schedule thus not specified.

\begin{theorem}\label{thm:main3}

	Suppose $G$ is a graph on $[n]$. Let ${\bf d}=\big(\Deg(1),\Deg(2),\dots, \Deg(n)\big)$ and ${\bf t, p}\in \mathbb{Z}^n $ where ${\bf p}\leq {\bf d}$.
	Then, in the
	\tp-system on $G$ with an arbitrary fair update schedule, the state ${\bf d}$ is reaching a stable state $C^{\bf t,p}=(C^{\bf t,p}_1,C^{\bf t,p}_2,\ldots, C^{\bf t,p}_n)$, and for any $i \in G$, 
	$$
	C^{\bf t,p}_i= C_{\bf t, p}(i).
	$$
	Equivalently, the ranks $C_{\bf t, p}(i)$ can be distributively computed via $C^{\bf t,p}_i$.
	Moreover, if ${\bf t}\leq {\bf t'}$ and ${\bf p} \leq {\bf p'}$, then $C^{\bf t,p}$ is reaching $C^{\bf t',p'}$
	in the $[{\bf t}', {\bf p}']$-system on $G$.
%	Moreover, we have for any $i \in G$, 
%	$$
%	C^{\bf t,p}_i= C_{\bf t, p}(i).
%	$$
\end{theorem}

Theorem~\ref{thm:main3} resonates with the claimed local computation of degree derivatives $C_{\bf t, p}(i)$
and its proof is given in the Methods section.
Depending on the deployment scenarios, there is possibly a simple central controller that has limited function, e.g., merely used to monitor the status of nodes and broadcast a signal for stopping the computation after all nodes have reached a steady state to save power.
%The proof of Theorem~\ref{thm:main3} is given in the section Methods.
According to Theorem~\ref{thm:main3}, the maximal \tp-separate chain if needed can be immediately constructed once the fixed point of the \tp-system is obtained.

% is expected to be complex if there is one.

\subsection{Long paths at a node}\label{sec5}

%Let $G=(V,E)$ be a graph.
The numbers $C_{\bf t, p}(i)$ encode lots of information about node $i$ including paths through $i$. A path of length $k-1$ is a sequence of distinct nodes $v_1, v_2,\ldots, v_k$ such that $v_i$ and $v_{i+1}$ ($1\leq i <k$) are neighbors in $G$.
The nodes $v_1$ and $v_k$ are called the terminals of the path.

It can be proven that if $p_v\leq -\deg(v)$, then $C_{\bf t, p}(v)$ depends on $t_v$ alone (see the SI).
For simplicity, we only utilize the numbers $C_{\bf t, p}(v)$ for $t_v$ being
a nonpositive constant $t$ and $p_v\leq -\deg(v)$ towards deriving path information,
although better performance may be achieved by exploring additional degree of freedom.
We also simply write $C_{\bf t, p}(v)$ as $C_{ t}(v)$.
Let
$$
\lambda(G)= \max \{ |\Deg(u)- \Deg(v)|: \mbox{$u$ and $v$ are adjacent in $G$} \}.
$$
We can show that for any $t \leq - \lambda(G)$ and any node $v$, we have
$C_t(v)=\Deg(v)$ (See the SI).

%%%
%%%%%%%%%%%%%%%%%%%%%%%%%%%%%%%%%%%%%%%%%%%%%%%%%%%%%%%%%%%%%%%%%%%%%%%%%%%%%%%%%%%%%%%%%%%%%%%%%%%%%%%%%%%%%%%%%%
%%%
\begin{figure}[!htb]
	\centering
	\includegraphics[scale=.8]{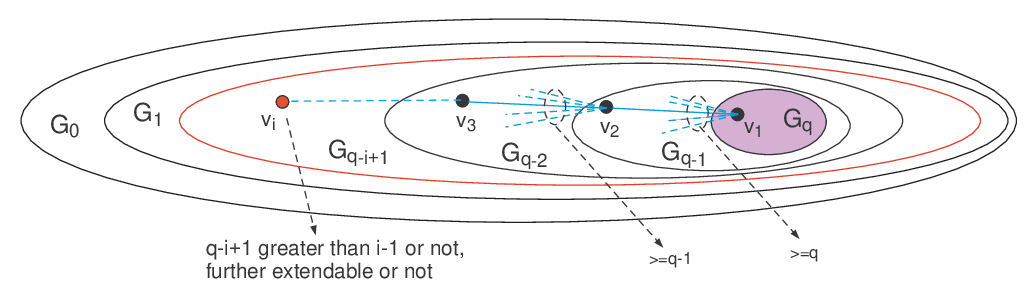}
	\caption{The idea for inferring a path in the chain for $t=-1$.}\label{fig:idea-illu}
\end{figure}
%%%
%%%%%%%%%%%%%%%%%%%%%%%%%%%%%%%%%%%%%%%%%%%%%%%%%%%%%%%%%%%%%%%%%%%%%%%%%%%%%%%%%%%%%%%%%%%%%%%%%%%%%%%%%%%%%%%%%%
%%%

%Next, we focus on the cases $t<0$.
%Given a graph $G$ and a vertex $v\in G$, 
The idea for inferring a path based on the ranks is illustrated in Figure~\ref{fig:idea-illu}.
To make it precise, for $t<0$, we additionally introduce the following notations:
\begin{align*}
	q_t(z) =\left \lfloor \frac{z}{|t|} \right \rfloor, \quad 0\leq r_t(z)=z \bmod |t| <|t|, 
	\quad \pi_{t,j}(z) =r_t(z)-t \cdot j \, ,
\end{align*}
%and $\xi_t(z)=\min \{ y: \pi_{t,y}(z)>1 \land y\geq 0 \}$ when such $y$ exist.
%\begin{align*}
%q_t(v) =\left \lfloor \frac{C_t(v)}{|t|} \right \rfloor, \quad 0\leq r_t(v)=C_t(v) \bmod |t| <|t|, 
% \quad \pi_{t,j}(v) =r_t(v)-t \cdot j \, .
%\end{align*}
%Let
%$$
%\xi_t(v)=\min \{ y: \pi_{t,y}>1\}.
%$$
Obviously, 
$$
z=r_t(z)-t\cdot q_t(z)=\pi_{t, q_t(z)}(z).
$$
In the case of $z=C_t(v)$, we simply write $q_t(C_t(v))$ as $q_t(v)$, and other notations are simplified analogously.
Let $N_G(v)$ denote the set of neighbors of $v$ in $G$ and $\overline{N}_G(v)=N_G(v)\cup \{v\}$. 
When the graph $G$, vertex $v$ and $t$ in question are clear from the context, $G$, $v$ and $t$ may be dropped in these quantities.
We assume no isolated nodes exist in $G$ discussed in this paper.
Let 
$$
J_t(z,x) =\min \Big\{q_t(z),  \left\lfloor \frac{q_t(z)+x-r_t(z)}{1-t} \right\rfloor  \Big\}.
$$
It is not difficult to check that $J_t(z,x)=0$ if $z\leq x$.
For $z=C_t(v)$, $J_t(v,x)$ is used to stand for $J_t(z,x)$, and we additionally use the short-hand $J_t(v)$
for $J_t(v,0)$.

\begin{theorem}[Lower bound]\label{thm:path}
	Let $G$ be a graph, and for any $v\in G$, let
	\begin{align*}
		L_{e}(v) &= \max_{-\lambda(G) \leq t \leq 0} \big\{ q_t(v|_t)  -J_t(v|_t) \big\},
	\end{align*}
	where $v|_t$ is any vertx in $\overline{N}(v)$ such that $C_t(v|_t)=\max\{C_t(w): w\in \overline{N}(v)\}$,
	and
	for any $u$, $q_0(u)-J_0(u)=C_0(u)$.
	Then, there exists a path of length at least $L_{e}(v)$ that has $v$ as a terminal.

\end{theorem}

\begin{theorem}[An improvement]\label{thm:path-t-ext}
	Let $G$ be a graph and $v\in G$. 
	%Suppose $\xi_t(v)$ exists and $J_t(v,0)\geq \xi_t(v)$.
	Then, there exists a path having $v$ as a terminal with length at least 
	$$
	\widehat{L}_{e}(v)= \max_{-\lambda(G) \leq t \leq 0} \left\lceil \frac{C_t(v|_t)(1-t)+t}{1-t+t^2}\right\rceil .
	$$
%	where $v|_t$ is any vertex in $\overline{N}(v)$ such that $C_t(v|_t)=\max\{C_t(u): u\in \overline{N}(v)\}$. 
	
\end{theorem}

Although our primary interest in this paper is concerned with
long paths starting with a vertex, we can analogously provide some lower bounds for the length of the longest
paths containing a prescribed node as well, for instance, the one below.

\begin{theorem}\label{thm:mid-path}
	
	Let $G$ be a graph and $v\in G$, and let
\begin{align*}
	L_{m}(v) = \max_{-\lambda(G) \leq t < 0} \left\{ [q_t(v)  -J_t(v)]+
		[q_t(v\dagger_t)-J_t(v\dagger_t,\, q_t(v)  -J_t(v) )]  \right\},
\end{align*}
	where $v\dagger_t$ is any vertex
	in ${N}(v)$ such that $C_t(v\dagger_t)=\max\{C_t(w): w\in N(v)\}$, 
	Then, there exists a path of length at least $L_{m}(v)$ that contains $v$.

\end{theorem}

%\begin{remark}
Proofs of the above theorems are provided in the Methods and Supplementary Information.
\emph{It has to be emphasized that these computations can be carried out
by the node $v$ itself.}
Therefore, our lower bounds above may be the first nontrivial estimates
on the length of the longest paths involving any specific individual node and utilizing only local information.
In addition, at the cost of the elegance of a uniform expression, we may improve our lower bounds
by distinguishing cases taking if $\Deg(v)>C_t(v)$, if $C_t(v)>C_t(v|_t)$, or alike into consideration.
See the application of these results to the graph in Figure~\ref{fig:exam} in the Supplementary Information.

\subsection{Navigation strategy}

A strategy (Algorithm~\ref{path-improve-start}) for relaying a message with local decisions
can be derived from the proof of Theorem~\ref{thm:path}.
Suppose $q_0(v)-J_0(v,x)=\max\{C_0(v)-x, 0\}$, and
\begin{align*}
	L_{end}(v, x) &= \max_{-\lambda(G) \leq t\leq 0} \big\{ q_t(v)  -J_t(v, x) \big\}.
\end{align*}
Moreover, 
let $A_{end}(v, x)$ denote the set of $t$'s that achieve the maximum $L_{end}(v, x)$.
A proof of Theorem~\ref{thm:alg1} below can be found in the Methods section.

\begin{algorithm}
	\caption{Navigation process initiated at $v$}\label{path-improve-start}
	\begin{algorithmic}[1]
		%	\Procedure{MyProcedure}{}
		\STATE {\bf Input} $G=(V,E)$ and $v\in V$
		\STATE $i\gets 1, \, j \gets 1 $
		\STATE $v_1 \gets v, \, w \gets v$
		\STATE $W \gets \varnothing$
		%\BState 
		\WHILE{ $N(w) \nsubseteq W$} 
		\STATE pick $t\in A_{end}(w, i-1)$, and among the neighbors of $w$, pick $x$ such that $x\notin W$ and $C_t(x)$ is maximal
		\STATE $W \gets W \bigcup \{v_i\}$
		\STATE $i \gets i+1$
		\STATE $v_i \gets x, \, w \gets x$
		\ENDWHILE
		\STATE $W \gets W \bigcup \{v_i\}$
		\RETURN $v_i, v_{i-1}, \ldots, v_1$
		%		\EndProcedure
	\end{algorithmic}
\end{algorithm}

\begin{theorem}\label{thm:alg1}
	
	%With the notation in Theorem~\ref{thm:path},
	Algorithm~\ref{path-improve-start} produces a path starting with $v$ of length at least $$L_{e}^{\circ} (v)=\max_{-\lambda(G)\leq t \leq 0}\{q_t(v)-J_t(v)\}.$$
\end{theorem}

\emph{It is apparent that nodes can make their own computations and decisions as long as they have certain computing power and can communicate with their neighbors.} To be specific, suppose the message starting with $v$ has arrived at the current node $u$.
What $u$ needs to know in order for making its own optimal decision is the index $i$ indicating it is the $i$-th node in the relay and which of its neighbors are
already in the race via communication with its neighbors. 
Nodes do not need to know the global structural information of the network, and each node is even not aware of which non-neighbors are involved in the
navigation process.

In order to evaluate our relaying strategy, we compare it with two natural
candidate algorithms: the current node randomly adds an unused neighbor to the path (i.e., random path) and randomly adds an unused neighbor of maximal degree to the path (i.e., max-degree based random path).
Six representative real-world networks from disparate fields, five Barab\'{a}si-Albert type networks, and five Erd\H{o}s-R\'{e}nyi random networks are used for evaluation. The real-world networks are: Email~\cite{email}, USAir~\cite{usair}, Jazz~\cite{jazz}, PB~\cite{pb}, Router~\cite{router}, and Email2~\cite{email2}. In brief, Email is the e-mail interchanges between members of the Univeristy Rovira i Virgili (Tarragona), USAir is  the US air transportation network, Jazz records the collaborations between jazz musicians, PB is the network of US political blogs, Router is a symmetrized snapshot of the structure of the Internet at the level of autonomous systems, and Email2 is the e-mail interchanges between members of the Computer Sciences Department of University College London.

%described in detail in the Supplementary Information.

\begin{figure}[t]
	\setlength{\tabcolsep}{5pt}
	\centering
	%	\begin{table}
	%		\centering
	\begin{tabular}{p{0.32\textwidth}p{0.32\textwidth}p{0.32\textwidth}}
		\begin{subfigure}{0.32\textwidth}
		%	\caption{Email}
			\centering
			\includegraphics[width=1.0\linewidth]{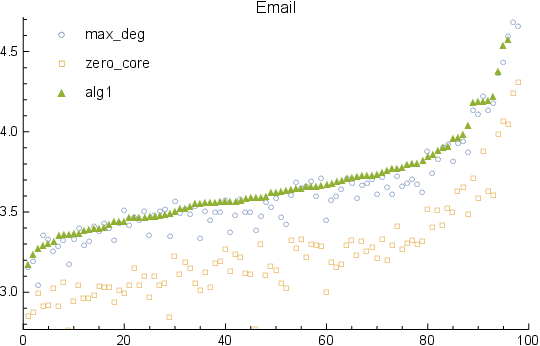}
		\end{subfigure}%
		&	\begin{subfigure}{0.32\textwidth}
		%	\caption{Email-dispersion}
			\centering
			\includegraphics[width=1.0\linewidth]{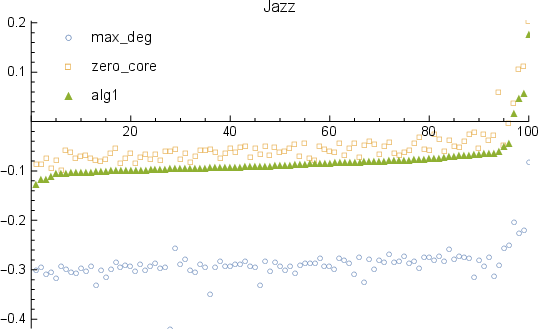}
		\end{subfigure}
		&\begin{subfigure}{0.32\textwidth}
		%	\caption{Jazz}
			\centering
			\includegraphics[width=1.0\linewidth]{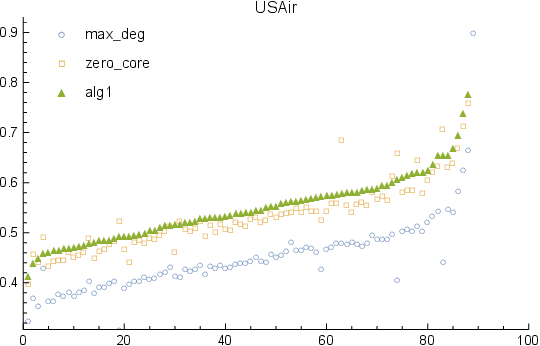}
			
		\end{subfigure}\\
		 	\begin{subfigure}{0.32\textwidth}
		%	\caption{Jazz-dispersion}
			\centering
			\includegraphics[width=1.0\linewidth]{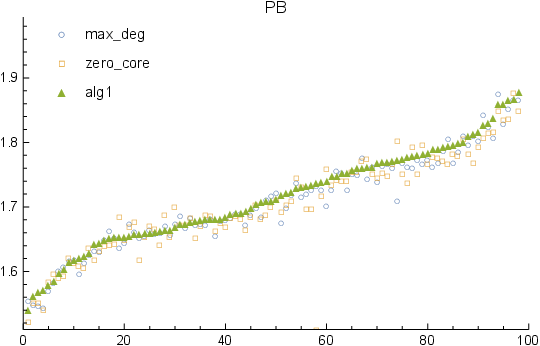}
			
		\end{subfigure}
		&\begin{subfigure}{0.32\textwidth}
		%	\caption{PB}
			\centering
			\includegraphics[width=1.0\linewidth]{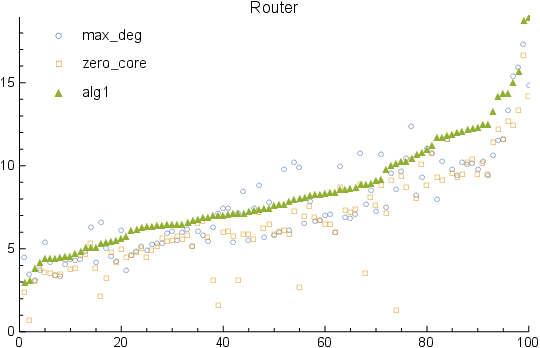}
			
		\end{subfigure}
		&	\begin{subfigure}{0.32\textwidth}
		%	\caption{PB-dispersion}
			\centering
			\includegraphics[width=1.0\linewidth]{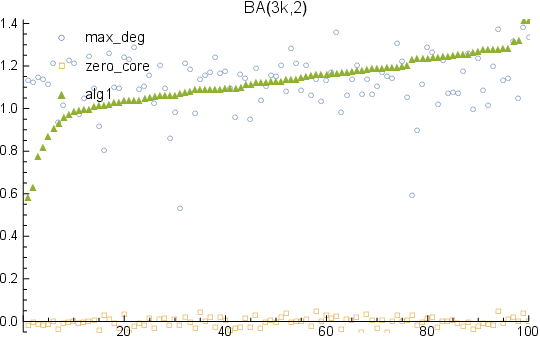}
			
		\end{subfigure}\\
	\begin{subfigure}{0.32\textwidth}
		%	\caption{Jazz-dispersion}
		\centering
		\includegraphics[width=1.0\linewidth]{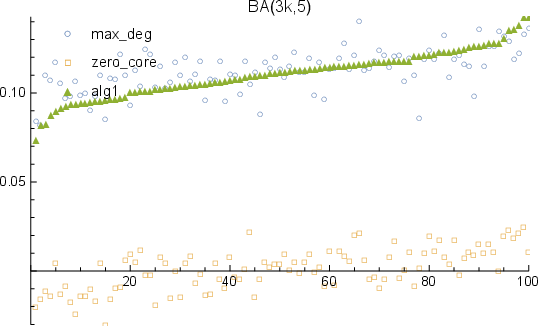}
		
	\end{subfigure}
	&\begin{subfigure}{0.32\textwidth}
		%	\caption{PB}
		\centering
		\includegraphics[width=1.0\linewidth]{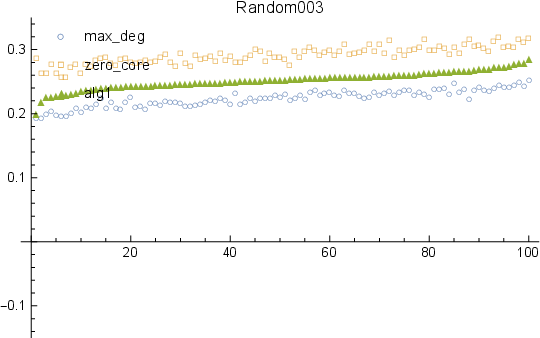}
		
	\end{subfigure}
	&	\begin{subfigure}{0.32\textwidth}
		%	\caption{PB-dispersion}
		\centering
		\includegraphics[width=1.0\linewidth]{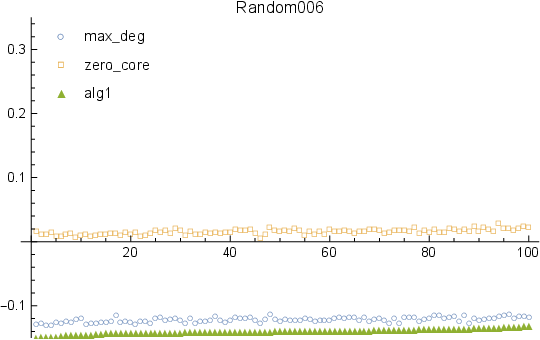}
		
	\end{subfigure}
	\end{tabular}
	%		\end{table}
	\caption{{\bf Algorithm~$1$ relays a message further}. {\normalfont For each network, algorithms are compared based on the found paths starting with each of $100$ random nodes (rearranged on the $x$-axis), where the $y$-coordinates show the normalized gains over the random path approach. Algorithm~$1$ (in green) is overall the best for networks that are not too dense.}}
	\label{fig:rand}
\end{figure}

For each network, we randomly pick 100 nodes as the message source, and each candidate algorithm is used to search 1000 paths starting from each picked node respectively.
Then, we compare the algorithms by comparing the corresponding average lengths of the paths. We actually compare two algorithms arising from our framework:
Algorithm~\ref{path-improve-start} and the one (i.e., zero-core) obtained by considering $t=0$ alone in Algorithm~\ref{path-improve-start}. 
At first, we use the random path length of a node $v$ as the benchmark and compute the ratio
$$
\frac{\mbox{path length of a candidate algorithm $-$ random path length}}{\mbox{random path length}}.
$$
As for real-world networks,
Figure~\ref{fig:rand} shows that except for Jazz network, the remaining algorithms are much better than the benchmark.
Our approach is generally better than the max-degree based approach, and Algorithm~\ref{path-improve-start} (more choices for $t$) is better than the zero-core ($t=0$ alone) approach as well.
In order to better illustrate this, we secondly employ the max-degree based approach as the benchmark
and compute the respective distributions of the normalized gains of our two algorithms. Figure~\ref{fig:max-deg} (in the SI) demonstrates that in all these networks, about $70\%$ of the time Algorithm~\ref{path-improve-start} is better (i.e., the gain is positive); For networks like USAir and Jazz, it is even $100\%$ better.

The Barab\'{a}si-Albert type network BA$(n,m)$ is a network of $n$ nodes where in the construction process each newly added node
is attached to $m$ existing nodes.
For these networks, as expected, the max-degree based relaying strategy is good, since a max-degree neighbor very likely has a neighbor of its own of high degree by the formation principle. However, even for these networks, our method still performs fairly well.

We have evaluated Erd\H{o}s-R\'{e}nyi random networks $G(n,p)$
for $p\in \{0.001,0.003,0.006,0.01,0.05\}$ and $n=3000$, named Random001, Random003, etc. It is observed that when $p$ is small (e.g., $p\in \{0.001,0.003\}$), random selection
is not a good strategy; however, when $p$ increases, i.e., the network becomes denser, the random selection approach is almost the best.
We believe the Jazz network is actually in the same situation, because it is the densest among the six real-world networks.
This trend can be also observed by comparing the gains over the random selection approach in the networks BA$(3k,l)$ for $2\leq l \leq 5$ that
the gains decrease as $l$ increases, i.e., the graphs become denser.
This is not a surprise, since there is a very long path for $p$ larger than the critical value $1/n$~\cite{aks81},
and in fact the depth-first searching algorithm can almost surely find a long path~\cite{dfs}.
Moreover, when $p$ is large enough (say $p>0.05$), there is almost no difference in performance for all strategies.

More figures are provided in the SI.
Overall, our method is good when applied to networks that are not very dense and most
practical networks are probably this case.

\section{Discussion}

The first nontrivial solutions to both research problems have been provided based on a single novel, rigorous mathematical framework here.
As we mentioned, we could not provide a theoretical performance justification of the estimated length at
a node right now.
One may think an analysis should be at least given to some special classes of graphs,
for instance, graphs with a lower bounded degrees or with connectivity characterization.
However, the notion of a class of graphs already implies certain global features of the graphs
which are not supposed to be known to a local node.
On the other hand, the problems studied are no easier than the Hamiltonian problem
or the travelling salesman problem.
Note that the vertices in the same graph are generally heterogeneous (the almost
homogeneous cases, e.g., $K_n$ or $K_{m,n}$, are not interesting for the studied problems).
It is thus difficult to imagine what a general theoretical statement may look like. 
%On the other hand, in a few networks, randomly picking an available neighbor may not be a bad relaying strategy but no node knows for sure about if the length of the longest paths starting with it is greater than two.
%We thus believe a theoretical, analytical justification is hard in view of the nature of the problem.
Of course, we can relax our assumption that certain global features like minimum degree that can be easily obtained are allowed to be
provided to the nodes (by a simple central controller), and integrating these global features into our method in order for
providing a potential better solution for our long-path problems will be investigated in future works.

Theorem~\ref{thm:path}, Theorem~\ref{thm:path-t-ext} and Algorithm~\ref{path-improve-start} rely on data $C_t(v)$ for all $-\lambda(G) \leq t\leq 0$ and $v \in G$.
Now we discuss the complexity of obtaining these data.
%In the worst case, we have to compute $C_t(v)$ for $-\lambda(G) \leq t \leq 0$.
We first compute the $C_{-\lambda(G)}$-values according to Theorem~\ref{thm:main3} with the initial system state being
the degree sequence ${\bf d}$.
For $t$ from $-\lambda(G)+1$ to $0$, we can obtain the $C_{t}$-values in accordance with Theorem~\ref{thm:main3} by searching the fixed point reached by the initial system state $(C_{t-1}(1),\ldots,C_{t-1}(n))$ in the $[t]$-system. Note that 
$$
{\bf 0} \leq C_0 \leq C_{-1} \leq \cdots \leq C_{-\lambda(G)} \leq {\bf d}.
$$
In addition, in each round of update of the vertices $v\in [n]$ (using the functions, $f_v$'s, with respect to the corresponding $t$), the state of at least one vertex will decrease unless we have already reached $C_t$. Since the total number of decrease steps from
${\bf d}$ to ${\bf 0}$ is at most
$\sum_i \Deg(i)$,
the time complexity is conservatively at most $n \cdot \sum_i \Deg(i)= 2n |E|$.
Thus, the worst-case time complexity is $O(n |E|)$.
Because our method is good for not very dense networks, the expected time complexity for networks where
our approach is utilized is probably at most $O(n^2)$.
As for computing the data $C_t(v)$, depending on the used scenario, a central controller with simple,
limited functions may be deployed. The controller does not have to maintain the adjacency relation of the very
network and implement complicated computation, and it only needs to send signal to ask an agent to update its status
and ask all agents to stop when a stable state is achieved or so. 
After the computation of $C_t(v)$ is completed, the remaining computation for estimating path length
and relaying decision at a node is simple.
Nevertheless, the detailed issues that need addressing in certain practical implementations of our
solutions are beyond the scope of the paper.

Here we have only considered \tp-separate chains where $t_v$ is a constant and $p_v$ is not relevant
for solving the long path problems.
Further optimization is possible by exploring additional degree of freedom of $\mathbf{t}$ and $\mathbf{p}$.
For instance, without $\mathbf{p}$, we always assume the rank will decrease from the present node (in $G_k$) to
its immediate successor (in $G_{k+t}$) in the current estimation. However, if $\mathbf{p}$ is in effect,
we know an immediate successor still in $G_k$ exists by construction, taking advantage of which may lead
to a better estimation. 
The restricted case is actually equivalent to the D-chain decomposition of networks introduced by Chen, Bura and Reidys~\cite{d-spec} where the induced D-spectra of nodes were used to effectively characterize the spreading power of nodes and the potential application to searching long paths was never noticed back there.
The underlying phylosophy of connecting certain subgraph chains of graphs
to dynamical systems is the same. However, such nice connections between the two fields
are not readily available. Hence,
we argue that \tp-separate chains are significant generalization of D-chains and an effort of careful construction (e.g., what kind of chains, and what update
functions may work) has to be made
to realize this generalization. As such, our \tp-separate chain framework is of independent interest and may be modified to adapt to directed networks and weighted networks, and can be applied to other network problems.

\section{Methods}

\subsection{Proof of Theorem~\ref{thm:main3}}
%\begin{proof}
Here we only prove $
C^{\bf t,p}_i= C_{\bf t, p}(i)$, the proof of the rest can be found in the Supplementary Material.
	Let ${\bf z}=\big( C_{{\bf t},{\bf p}}(1)\ldots, C_{{\bf t},{\bf p}}(n)\big)$.
	Then, in view of Proposition~\ref{main-thm-1} and~\ref{main-thm-2}, it suffices to prove that
	the state ${\bf z}$ is a fixed point of the $[{\bf t},{\bf p}]$-system
	and the state ${\bf d}$ is reaching ${\bf z}$.
	We shall first prove:
	
	{ \emph{Claim}~$1$.} The state ${\bf z}$ is a fixed point of the $[{\bf t},{\bf p}]$-system.

	Let $\mathcal{C}: G_0\geq G_1\geq \cdots $ be the maximal $[{\bf t}, {\bf p}]$-separate chain of $G$.
	First, suppose $C_{{\bf t},{\bf p}}(v)=i$ for a vertex $v$. By definition, $v$ belongs to $G_i$ but not $G_{i+1}$,
	which tells us:
	\begin{itemize}
		\item[ (i)] there exist at least $i$ neighbors of $v$ contained in $G_{j}$ ($j=\max\{0, i+t_v\}$) and at least $\max\{0,i+p_v\}$ neighbors of $v$ contained in $G_i$. Note that for any $u$ among these said neighbors in $G_j$, we have $C_{{\bf t},{\bf p}}(u)\geq i+t_v$, and for any $u'$ among these neighbors contained in $G_i$ we have $C_{{\bf t},{\bf p}}(u')\geq i$ by definition. Thus, among the neighbors of $v$, there exist at least $i$ of them with values at least $i+t_v$ and at least $\max\{0,i+p_v\}$ of them with values at least $i$ in ${\bf z}$. This leads to that $f_v({\bf z})\geq i=C_{{\bf t},{\bf p}}(v)$;
		\item[(ii)] it is impossible that $i+1$ neighbors of $v$ are contained in $G_{j'}$ ($j'=\max\{0, i+1+t_v\}$) and $i+1+p_v$ neighbors of $v$ are contained in $G_{i+1}$. Otherwise, the chain $G_0\geq \cdots \geq G_i\geq G_{i+1}\bigcup \{v\}\geq G_{i+2}\geq \cdots$ yields a $[{\bf t}, {\bf p}]$-separate  chain, which contradicts the maximality of $\mathcal{C}$. Hence, the case that at least $i+1$ of the neighbors of $v$ have values at least $i+1+t_v$ and at least $i+1+p_v$ of the neighbors with values at least $i+1$ in ${\bf z}$ cannot happen. This implies
		$f_v({\bf z})< i+1$.
	\end{itemize}
	From (i) and (ii), we conclude $f_v({\bf z})=i=C_{{\bf t},{\bf p}}(v)$ for any $v$. Therefore, ${\bf z}$ is a fixed point.
	
	We next shall show that ${\bf d}$ is reaching ${\bf z}$. If ${\bf z=d}$, we are done.
	Otherwise we clearly have ${\bf z<d}$.
	For this case, in the light of
	Proposition~\ref{main-thm-2} in Supplementary Information, it suffices to show that any state ${\bf y\leq d}$ such that ${\bf y>z}$ or y is uncomparable with ${\bf z}$ is not a fixed point.
	
	We prove by contradiction.
	Suppose ${\bf y}$ is such a state which is a fixed point. Then there must be a coordinate indexed by some $v$ that satisfies $y_v> C_{{\bf t},{\bf p}}(v)$. Consider the sequence of subgraphs induced by the sequence of sets of vertices $S_0 \supseteq S_1\supseteq \cdots \supseteq S_{m}$ for a sufficient large number $m$ (say $m>n+\max\{t_v: v\in V\}$) which are iteratively constructed as follows:
	\begin{itemize}
		\item[(1)] set $S_{y_v}=\{v\}$ and $S_r=\varnothing$ for $r\neq y_v$;
		\item[(2)] for $r$ from $0$ to $m$, if $u\in S_r$, then set
		\begin{align*}
			 S_{\max\{0,r+t_u\}}&=S_{\max\{0,r+t_u\}}\bigcup 
			\{w: \mbox{$w$ is a neighbor of $u$ and $y_w\geq r+t_u$}\},\\
			 S_{r}&=S_{r}\bigcup \{w: \mbox{$w$ is a neighbor of $u$ and $y_w\geq r$}\},
		\end{align*}
		and for $0\leq r \leq m$, set $S_r=\bigcup_{j=r}^m S_j$;
		\item[(3)] iterate (2) until the sequence  $S_0 \supseteq S_1\supseteq \cdots \supseteq S_{m}$ becomes stable, i.e., $S_r$ does not change for any $0\leq r \leq m$ when further executing (2).
	\end{itemize}
	Clearly, by construction we have $S_{r}\subseteq S_{r-1}$.
	By abuse of notation, we denote by $S_r$ the subgraph induced by $S_r$ as well. Then we have a chain of graphs $S_0\geq S_1\geq \cdots \geq S_{m}$.
	Let $\tilde{{\bf t}}$ be the restriction of ${\bf t}$ to that of the vertices in $S_0$ and $\tilde{{\bf p}}$ be the restriction of ${\bf p}$ to that of the vertices in $S_0$.
	We proceed to show that
	
	{\emph{Claim}~$2$.} The chain $S_0\geq S_1\geq \cdots \geq S_{m}$ gives a $[\tilde{{\bf t}},\tilde{{\bf p}}]$-separate chain of the graph $S_0\leq G$. That is, for any $u\in S_0$, if $u\in S_r$, then there are at least $r$ neighbors of $u$ contained in $S_{j}$ where $j=\max\{0, r+t_u\}$ and at least $\max\{0,r+p_u\}$ neighbors of $u$ contained in $S_r$.

	Since ${\bf y}$ is a fixed point by assumption, there exist at least $y_u$ neighbors of $u$
	whose corresponding values in ${\bf y}$ are at least $y_u+t_u$ and at least $y_u+p_u$ neighbors whose corresponding values in ${\bf y}$ are at least $y_u$.
	By construction of (2), if $u\in S_r$, then $r\leq y_u$.
	Furthermore, since $u$ has at least $y_u$ neighbors $w$ such that $y_w \geq y_u+t_u \geq r+t_u$,
	these vertices $w$ are contained in $S_{\max\{0, r+t_u\}}$. Analogously, there are at least $y_u+p_u\geq r+p_u$
	neighbors of $u$ contained in $S_r$.
	Accordingly, the chain $S_0\geq S_1\geq \cdots \geq S_{m}$ gives a $[\tilde{{\bf t}},\tilde{{\bf p}}]$-separate chain, whence Claim~$2$.

	In view of Claim~$2$, it can be easily checked that the chain 
	$
	G_0 \geq  S_1 \geq \cdots \geq S_{y_v} \geq \cdots
	$
	gives a $[{\bf t,p}]$-separate chain of $G$. Therefore, we have $C_{\bf t,p}(v)\geq y_v$ since $v\in S_{y_v}$, which yields a contradiction.
	Hence, ${\bf y}$ cannot be a fixed point.
	According to Proposition~\ref{main-thm-2}, ${\bf d}$ cannot reach a fixed point that is smaller than ${\bf z}$ either.
	Therefore, ${\bf d}$ is reaching ${\bf z}$, and the proof follows.
%\end{proof}

	\subsection{Proof of Theorem~\ref{thm:path}}

	%\begin{proof} 
The derived path lengths from the situation $t=0$ are separately proved in the Supplementary Information.
	We focus on the situation $t<0$ below.
	 It can be checked that the following holds:
	\begin{align*}
		J_t(v) &=\max\{ j \in \mathbb{Z}\mid \pi_{t,j}(v)
		\leq q_t(v)-j \}  ,\\
		q_t(v)-J_t(v) & \geq q_t(u)-J_t(u), \quad \mbox{if $C_t(v)\geq C_t(u)$}.
	\end{align*}
	Suppose $\mathcal{C}: G_0 \geq G_1 \geq \cdots \geq G_m$ is the maximal $t$-separate chain of $G$. By construction, for $k\leq m$, $G_k\neq \varnothing$.
	Since the minimum degree $\delta(G)>0$, i.e., no isolated nodes, we also have $G_0=G_1=G$ implying $C_t(u)=\pi_{t, q_t(u)}(u) \geq 1$ for any $u\in G$. As a result, we have for any $u$,
	$$
	\pi_{t, q_t(u)}(u)\geq 1>q_t(u)-q_t(u)=0.
	$$
	Thus, $J_t(u) \leq q_t(u)-1$.

	\emph{Claim~$1$.} For a fixed $t<0$, there exists a path in $G$ of length at least $q_t(v)  -J_t(v)$ which has $v$ as a terminal. \\
	Start with $v_q=v \in G_{\pi_{t,q_t}}$, and pick any one of its neighbors, say $v_{q-1}$, contained in $G_{\pi_{t,(q_t-1)}}$, and obtain a path of length one as a consequence.
	If $J_t=q_t-1$, then we are done.
	Suppose $J_t<q_t-1$,
	and suppose $v_j \in G_{\pi_{t,j}}$, $J_t+1\leq j \leq q_t-1$, has been picked.
	Since $j\geq J_t+1$, we have 
	$$
	\pi_{t,j}>q_t-j\geq 1.
	$$
	Note that for $v_j \in G_{\pi_{t,j}}$ we automatically have $\Deg(v_i)\geq \pi_{t,j}$.
	This implies that there exists at least one neighbor of $v_j$ that is not on the path $\mathfrak{L}:  v_q, v_{q-1},\ldots, v_{j+1}$.
	Accordingly, we can add one more vertex $v_{j-1}$ that is a neighbor of $v_j$ to the path.
	Hence, we eventually obtain a path of length $q_t-J_t$ having $v$ as a terminal,
	and Claim~$1$ follows. 
	
		\emph{Claim~$2$.} For a fixed $t<0$ and $u\in N(v)$, there exists a path of length at least $q_t(u)  -J_t(u)$ which has $v$ as a terminal in $G$. \\
	First we can analogously obtain a path of length $q_t(u)  -J_t(u)$ starting with $u$: $u=u_q, u_{q-1},\ldots, u_{J_t(u)}$.
	Note that we have more than one choice for $u_i$ for any $q_t(u)-1\geq i \geq J_t(u)+1$ by construction so that it can be guaranteed that
	$u_i\neq v$ for any $q_t(u)\geq i \geq J_t(u)+1$.
	Next, if $u_{J_t(u)}=v$, we are done. Otherwise, we have the path
	$v, u, u_{q-1},\ldots, u_{J_t(u)}$. In either case, Claim~$2$ follows.
	
	Using the fact that $q_t(v|_t)-J_t(v|_t) \geq q_t(v)-J_t(v)$ and
	taking the maximum over all $t$'s, $-\lambda(G) \leq t \leq 0$, we conclude that 
	there exists a path of length at least $L_{e}(v)$ that has $v$ as a terminal.

\subsection{Proof of Theorem~\ref{thm:alg1}}

First for $i=1$ and $v_1=v$, it is obvious that $L_{end}(v_1,0)=L_{e}^{\circ}(v)$.
Suppose next $t_1\in A_{end}(v_1,0)$ and $v_2$ are picked. By construction, if in the following steps, the selected $t$'s are always $t_1$, then it is clear that the length of the obtained path is larger than $L_{e}^{\circ}(v)$. Otherwise, it suffices to show that for a different $t_2$ after the first step, the length of the obtained path starting from $v_2$ is larger than $L_{e}^{\circ}(v)-1$. To that end, for $v_2$, we first have $C_{t_1}(v_2)\geq C_{t_1}(v_1)+t$ and $q_{t_1}(v_2)-J_{t_1}(v_2,1)\geq L_{e}^{\circ}(v)-1$ from the proof of Theorem~\ref{thm:path}. Thus, 
$$
L_{end}(v_2,1)\geq q_{t_1}(v_2)-J_{t_1}(v_2,1) \geq L_{e}^{\circ}(v)-1.
$$
Next, if $t=t_2 \in A_{end}(v_2,1)$ is picked and does not change in the following steps, then the obtained path starting from $v_2$ is larger than $L_{e}^{\circ}(v)-1$. If a different $t$ is picked later, then we can iterate analogous argument and conclude that the length of the obtained path is always larger than $L_{e}^{\circ}(v)$.

\subsection{Benchmark methods}

The used benchmarks for performance analyses are based on two natural approaches since no other nontrivial approaches are known to us: randomly adding unused neighbor and randomly adding unused neightbor of maximal degree.
As for the former, the current node scans its neighbors and uniformly randomly picks a neighbor which is not on the so-far obtained path.
Analogously, for the latter approach, the current node scans the degree of its neighbors not on the so-far obtained path and then uniformly randomly picks one node among those of maximal degree.

\section*{Acknowledgments}

The author was supported by the Anhui Provincial Natural Science Foundation of China\\ (No.~2208085MA02)
and Overseas Returnee Support Project on Innovation and Entrepreneurship of Anhui Province (No.~11190-46252022001).

\section*{Declarations}
\noindent{\bf Data Availability: }Available upon request.

\noindent{\bf Conflict of Interest:} None.

%\end{appendix}

\newpage

\begin{appendix}

\begin{center}	
{\LARGE \bf Supplementary Information}
\end{center}

\section{Emergence of \tp-chains}

 D-chains (dendrite chains) of networks were introduced by Chen, Bura and Reidys~\cite{d-spec}.
 The finding that the maximal D-chains can be computed by a dynamical system was really exciting.
 The author later realized that there is still a drawback in the formulation of D-chains for the
 purpose of characterizing substructures in networks. That is,
 in the formulation of a D-chain of order $t<0$, $G_0 \geq G_1 \geq \cdots \geq G_m$,
 it is nice that the reachability of a subgraph $G_i$ to the outside of itself (i.e., $G_{i+t}$)
 has been characterized, but there is a lack of characterization of the inner structure $G_i$ itself.
 It could be that $G_i$ is just an independant set.
 However, in some applications, like identifying a super-spreading subset of nodes in the case of information or disease
 spreading, heuristically a subset of nodes which are well inter-connected themselves and at the same time well connected to
 the rest of nodes in the network may be a good candidate set.
 This was the motivation of extending the D-chain framework by introducing another layer of conditions.
 However, it is more than just arbitrarily introducing additional conditions since
 it is useless if there is no easy way to compute the generalized chains.
Efforts have to be put into careful constructions, and eventually \tp-chains emerge.

\begin{figure}[!htb]
	\centering
	\includegraphics[width=0.8\textwidth]{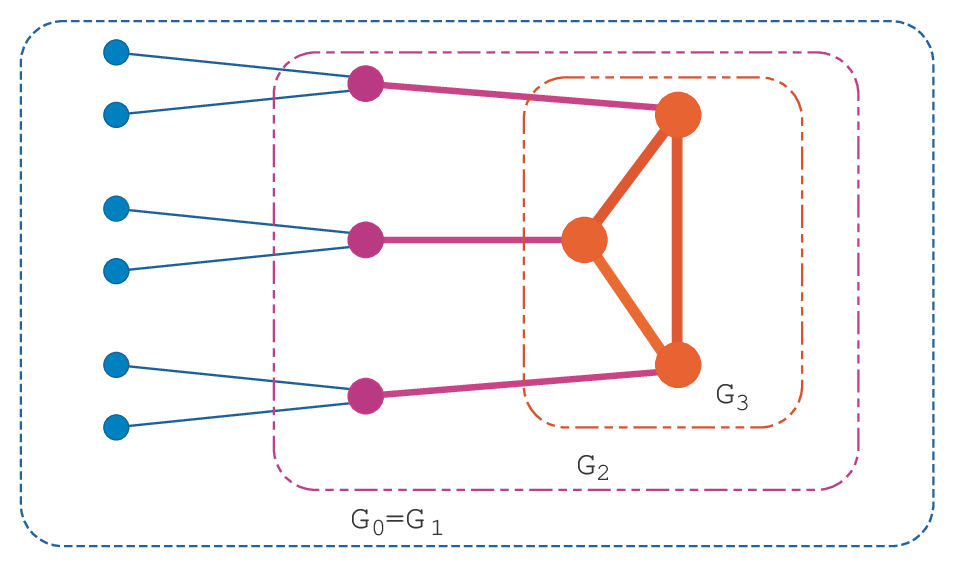}
	\caption{A \tp-separate chain $G_0 \geq G_1 \geq G_2 \geq G_3$ for a simple network $G$ of $12$ nodes and $12$ edges where $t_v=-1$ and $p_v=-1$ for any node $v$, i.e., a node is contained in $G_i$ if the node has at least $i-1$ neighbors in $G_i$ and at least $i$ neighbors contained in $G_j$ with $j=\max\{0, i-1\}$.}
	\label{fig:tp-decomposition}
\end{figure}

An example of a \tp-separate chain decomposition of a graph is presented in Figure~\ref{fig:tp-decomposition}.

In the definition of a \tp-separate chain, if the requirement specified by ${\bf p}$ is removed and $t_v$ is limited to a non-positive constant $t$ for any $v\in V$,
then the corresponding chain is a D-chain of order $t$.
Thus, \tp-separate chains greatly generalize D-chains in the following aspects.
First, in such a generalized chain, the parameter $t$ is vertex specific instead of a common constant for all vertices, which is used to control the corresponding subgraph $G_j$. Secondly, for each vertex $v$, one additional parameter $p_v$ is designed to control the number of neighbors of $v$ contained in $G_i$ itself if $v \in G_i$.

Technical side,
at first, it should be noted that it is crucial to allow empty graph $\varnothing$ in the chain which
was not relevant for D-chains~\cite{d-spec}.
In the empty graph, there is no vertex. The number of neighbors of an (outside) vertex with respect to the empty graph is zero.
Secondly, D-chains of order $t$ of a graph $G$ always exist. But, for some ${\bf t}, {\bf p}\in \mathbb{Z}^n$, there may be no \tp-separate chains for $G$ at all.
For instance, if $p_v>n$ for any vertex $v$ of $G$, then \tp-separate chains do not exist. Since the degree
of any vertex $v$ in any subgraph of $G$ is smaller than the total number
of vertices $n< \max\{0, i+p_v\}$ (for any $i\geq 0$).
Thirdly, regarding computation, it is not immediately clear that a dynamical system approach still exists and which local functions work.

Note that given any \tp-separate chain $G_0\geq  G_1  \geq  \cdots \geq G_m$, the chain $G_0\geq  G_1  \geq  \cdots \geq G_m\geq  \varnothing \geq \cdots\geq \varnothing$
is also a \tp-separate chain. Thus, given any \tp-separate chain, there are many chains essentially the same as the given one.
In other words, some $\varnothing$'s are essentially necessary in order to rigorously satisfy the degree condition, some are not.

For a vertex $v$ of $G$, we denote by $\Deg_G(v)$ the degree of $v$ in $G$,
and we write $\Deg(v)$ for short if the graph $G$ referred to is clear from the context.
The following proposition gives a sufficient and necessary condition for the existence of \tp-separate chains.

\begin{proposition}\label{prop:existance}
	Let $G=(V, E)$ and ${\bf t}, {\bf p}\in \mathbb{Z}^n$.
	Then, there exists a \tp-separate chain of $G$ if and only if $p_v \leq \Deg(v)$ for any $v\in V$.
\end{proposition}
\begin{proof}
	Consider the chain of size $0$, i.e., $G_0=G, \varnothing, \varnothing, \ldots $ where there are a sufficient number of $\varnothing$'s. If $p_v\leq \Deg(v)$, then $0 \leq \max\{0, 0+p_v\} \leq \Deg(v)$. Then it is clear that for any $v\in G_0$, there are at least zero neighbors of $v$ in $G_j$ for any $j\geq 0$ (in particular for $j= \max \{0, 0+t_v\}$) and at least $\max\{0, 0+p_v\}$ neighbors of $v$ in $G_0$.
	Since $G_i=\varnothing$ for $i>0$, there is nothing to check and the degree condition is automatically satisfied. Thus the chain is a \tp-separate chain.
	
	Conversely, suppose $G_0=G, G_1, \ldots, G_m$ is a \tp-separate chain. Suppose vertex $v$ is contained in $G_i$ for some $i\geq 0$.
	Then, by definition, there must be at least $i+p_v$ neighbors of $v$ in $G_i$. Since $G_i \leq G$, we have $i+p_v \leq \Deg(v)$, which implies $p_v \leq \Deg(v)$. This completes the proof.
\end{proof}

\section{Monotone-contractive systems}

In a dynamical system where a node can take a state from the set $P$,
suppose there is a linear order `$\leq $' on $P$. Let $P^q=\{(x_1,x_2,\dots, x_q): x_j \in P,\, 1\leq j \leq q\}$. Then, 
there is a natural partial order on $P^q$ as follows:  $(x_1,x_2,\dots , x_q)\leq (y_1,y_2,\dots, y_q)$ if for all $1\leq j \leq q$, $x_j\leq y_j$ in $P$.
A function $g: P^q\rightarrow P$ is called monotone, if for any ${\bf x\leq y}$ in $P^q$, we have $g(x)\leq g(y)$ in $P$.
Fox example, the binary functions `AND' and `OR' on $P^q=\{0,1\}^q$ are monotone.

A local function $f_{i}: (x_i, x_{k_1},x_{k_2},\dots, x_{k_i})\mapsto x'_i$ is called contractive (where
$k_1,\ldots, k_i$ are the neighbors of $i$), if
for any argument $(x_i, x_{k_1},x_{k_2},\dots, x_{k_i})\in P^{k_i+1}$, $x'_i \leq  x_i$ holds.
It is easy to check that if $f_{i}$ is the binary function `AND', then it is contractive under the assumption $0<1$.
A dynamical system where every local function is monotone and contractive is called a monotone-contractive (M-C) system.

The following results on M-C systems are relevant.

\begin{proposition}[\cite{d-spec}]\label{main-thm-1}
	For any two fair update schedules $W$ and $W'$, a system state ${\bf x} \in P^n$ is reaching the same fixed point ${\bf z}\leq {\bf x}$ under the two M-C systems
	$[G,f,W]$ and $[G,f,W']$.
	Furthermore, any system state ${\bf y} \in P^n$ such that ${\bf z}\leq {\bf y} \leq {\bf x}$ is reaching the fixed point ${\bf z}$.
\end{proposition}

As a consequence of Proposition~\ref{main-thm-1}, the exact form of the update schedule does not matter when come to discussing fixed points in M-C systems. Therefore,
we shall not explicitly specify the update schedules of M-C systems unless it is necessary.

\begin{proposition}[\cite{d-spec}]\label{main-thm-2}
	Suppose the system state ${\bf x} \in P^n$ is reaching the fixed point ${\bf z} \leq {\bf x}$ under an M-C system. Then,
	any state ${\bf y} \in P^n$ such that ${\bf z} \leq {\bf y} \leq {\bf x}$  or ${\bf y}< {\bf x}$ but not comparable to ${\bf z}$ is not a fixed point of the M-C system.
\end{proposition}

\begin{figure}[!htb]
	\centering
	\includegraphics[width=0.6\textwidth]{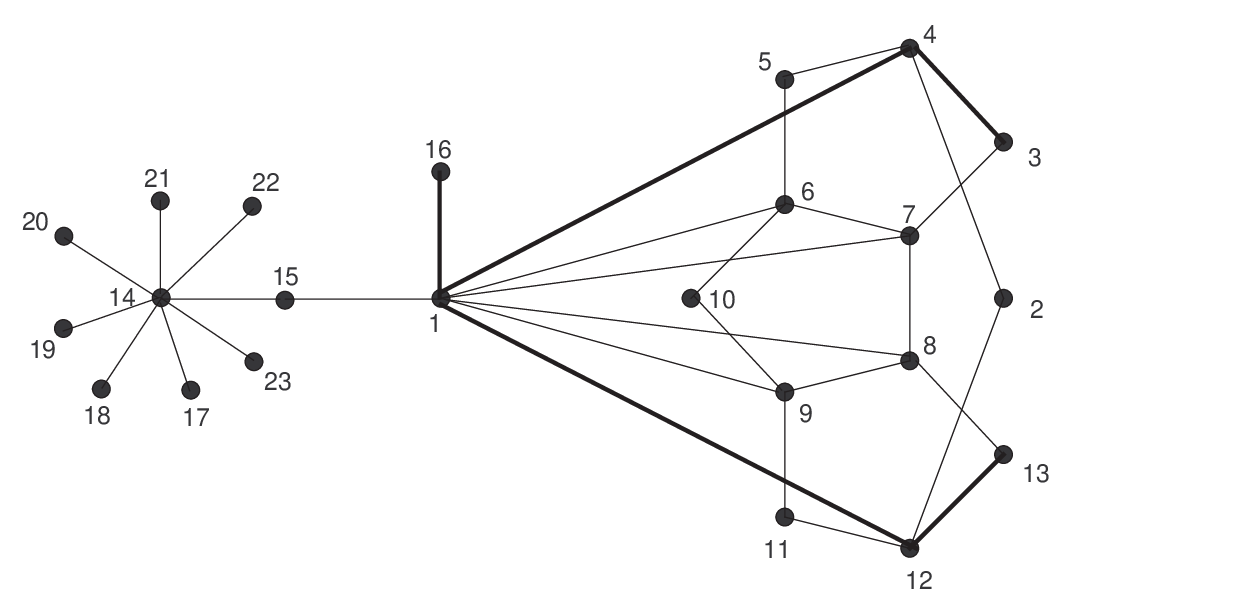}
	\caption{A graph $G$.}
	\label{fig:exam}
\end{figure}

%\begin{example}
Let us consider the graph $ G$ in Figure~\ref{fig:exam}, where the vertex set is $V=[23]$,
the number of edges is $31$, and the minimum node degree is $1$.
Certainly, an individual node does not have to know any of these global information in our method.

For $t=0$, the maximal chain is given by
$$
G_0=[23], \, G_1=[23], \, G_2=[13],
$$
where we only indicated the vertex set of $G_i$.
From Proposition~\ref{prop:core-bound}, we can conclude that there is a path of length at least two
starting with and containing node $16$,
e.g.,~$\mathcal{P}_0: 16,1,4$ or $\mathcal{P}'_0: 16,1,12$.

For $t=-1$, the maximal chain is given by
\begin{align*}
	G_0=[23], \, G_1=[23], \, G_2=[15], \, 
	G_3=\{1,4,6,7,8,9,12\}, \, G_4=\{1\}.
\end{align*}
Accordingly, since $C_{-1}(1)=4$, we compute $\left\lceil \frac{4(1+1)-1}{1+1+1}\right\rceil=3$ and we conclude from Theorem~\ref{thm:path-t-ext} that there is a path of length at least $3$ in $G$ that starts with $16$, e.g.,~$\mathcal{P}_{-1}: 16,1,4,3$,
better than the approximation from the case $t=0$.
Actually, we can further optimize our lower bounds by distinguishing distinct cases.
For example, here we know that $\Deg(16)=1$, so it cannot be contained in the initially constructed
path $\mathfrak{L}$ starting with $1$. Otherwise, $\Deg(16)\geq 2$. However, the length of $\mathfrak{L}$
is at least $3$, say $1,4,3,7$. Thus, by adding the edge $16-1$ to $\mathfrak{L}$, we have a path of length $4$ that starts
with $16$: $16,1,4,3,7$.

In terms of the ``one-way" relaying algorithm, suppose we start with node $15$.
If we randomly pick a neighbor of maximal degree, we may either pick $14$ or $1$.
In case that $14$ is picked, the length of the resulting relaying path has only length two.
However, applying our proposed algorithm, node $1$ will be inevitablly chosen and the resulting relaying path will be much longer.

Finally, all networks used for evaluation are treated as undirected, and their basic topological features are shown in Table~\ref{table:1}.

\begin{table}[h!]
	\centering 
	\setlength{\tabcolsep}{7mm}
	%		\begin{adjustbox}{width=\textwidth}
	\begin{tabular}{ |c|c | c |c |c| c|  }
		\hline

		network name & $N$ & $N_E$& $k_{max}$ & $<k>$ & $\lambda$\\
		\hline
		%		\rowcolor{lightgray}
		Email &1,133 &5,451 &71 &9.62 &70 \\
		%\rowcolor{lightgray}
		Jazz  &198 &742 &100 &27.69 &99 \\
		%	\rowcolor{lightgray}
		PB    &1,222 &16,714 &351 &27.35 &350 \\
		Router &5,022 &6,258 &106 &2.49 &105 \\
		%	\rowcolor{lightgray}
		USAir &332 &2,126 &139 &12.80 &138 \\
		Email2 &12,625 &20,362 &576 &3.22566 &575\\
		
		\hline
	\end{tabular}
	%	\end{adjustbox}
	\caption{$N$ denotes the number of nodes, $N_E$ denotes the number of edges, $k_{max}$ the maximum degree, $<k>$ the average degree.}
	\label{table:1}
\end{table}

\end{appendix}

\end{document}